\documentclass[12pt]{article}
\begin{document}
\title{Recovering the M-channel  Sturm-Liouville operator from M+1 spectra.}

\author{V. M. Chabanov  \\
Laboratory of theoretical physics, JINR, \\
Dubna, 141980, Russia \\
email: chabanov@thsun1.jinr.ru \\
 } \date{}

\maketitle

\begin{abstract}
For a system of M coupled Schr\"odinger equations, the
relationship is found between the vector-valued norming constants
and  $M+1$ spectra corresponding to the same potential matrix but
different boundary conditions. Under a special choice of
particular boundary conditions, this equation for norming vectors
has a unique solution. The double set of norming vectors and
associated spectrum of one of the $M+1$ boundary value problems
uniquely specifies  the matrix of potentials in the multichannel
Schr\"odinger equation.
\end{abstract}
PACS
%03.65.-w,
02.30.Zz

\section{Introduction}

Consider the  system of coupled one-dimensional Schr\"odinger
equations
\begin{eqnarray}
-\frac{d^2}{dx^2} \Psi _{\alpha}(x) + \sum_{\beta} V_{\alpha \beta
}(x) \Psi _{\beta} (x) = (E-\varepsilon_{\alpha }) \Psi_{\alpha
}(x), \enskip \alpha = 1,...,M. \label{schr}
\end{eqnarray}
In this system, each equation is referred to as a `channel' and
$\varepsilon_{\alpha }$'s are the energies of channel
`thresholds'. Once $E \ge \varepsilon_{\alpha }$, it is said that
$\alpha $'s threshold becomes open. The system (\ref{schr}) is a
matrix generalization of the ordinary one-dimensional
Schr\"odinger equation. The coupled Schr\"odinger equations
originate in the Feshbach's unified theory \cite{F} of nuclear
reactions and correspond to so-called approximation of the strong
coupling (when a finite number of equations in (\ref{schr}) is
left). Now, that method, renewed and generalized (see, e.g.
\cite{ZS}), finds a lot of applications and, rightfully, is one of
the most universal tools for microscopic description of systems
with many degrees of freedom (nuclear structure, reactions,
molecules, etc).

The inverse problem for multichannel Schr\"odinger equation
(\ref{schr}) has also been developed [3-5]. As in one-channel
case, one can uniquely restore the potential matrix $V_{\alpha
\beta }(x)$ from the spectral measure that, e.g., for the case of
bounded interval, is specified by the complete set of eigenvalues
$E_{n}$ and so-called norming vectors (spectral weight vectors)
$\gamma _{\alpha }(E_{n})$. These vectors characterize the
behaviour of the normalized wave functions $\Psi
_{\alpha}(x,E_{n})$ at one of the boundaries of interval (or at
the origin for a half-axis problem, etc.), see also below.

At the same time, in the one-channel case we have more variants of
the inverse problem. Among them, there is a statement of inverse
eigenvalue problem on a bounded interval where no norming
constants occur. Namely, the potential is uniquely recovered from
a knowledge of only two different spectra, each for a distinct
pair of homogeneous boundary conditions (with the same potential)
\cite{Borg} . There were established  necessary and sufficient
conditions of the solvability of the  inverse Sturm-Liouville
problem from two-spectra, see, e.g., the book \cite{Lev}.

Till now, one attempt to generalize this theorem to the
multichannel Sturm-Liouville operator has been known  to the
author -- the article \cite{Zak} (the case of a finite-difference
operator). Though not complete, this work gave an idea of the
existence of such a generalization in principle. No doubt, the
possibility of deriving potential matrix from a certain set of
spectra would contribute to the multichannel inverse problem
theory. In present article, results concerning that problem are
obtained. It is found that $M+1$ spectra determine the $V_{\alpha
\beta }(x)$ and, under special conditions, it is possible to
uniquely restore multichannel Sturm-Liouville operator.

The central idea of the paper is to derive the relationship
between $M+1$ spectra and M-component norming vector $\gamma
_{\alpha }(E_{n})$ associated with one of the $M+1$ boundary value
problems. Then, having the double set of eigenvalues and norming
vectors, one can uniquely restore an interaction matrix (by
Gel'fand-Levitan procedure).

Next section is devoted to setting forth these results. We shall
find the sought expression which, however, does not guarantee the
uniqueness in itself. Only under a special choice of boundary
conditions it can be represented in a form of system of linear
algebraic equations which give a simple criterion of the
uniqueness and solvability. For the sake of the reader's
convenience, the narrative is organized so that it goes partially
in parallel with standard derivation of two spectra formulas given
in \cite{Lev}, chapter 3.

\section{Derivation of the formula for norming vector}

We are beginning this section with preliminary notations.
 Let us rewrite the system (\ref{schr}) in a more symbolic form
as follows
\begin{eqnarray}
-\frac{d^2}{dx^2} y(x) + {\hat V}(x) y(x) = \lambda y(x), \enskip
x \in [0, a] \label{schrsymb}
\end{eqnarray}
where $y$ stands for the whole vector-column solution
\begin{eqnarray}
y(x) \equiv \left( \begin{array}{c} \Psi_{1}(x)\\
. \\
.\\
 .\\
\Psi_{M}(x)\end{array} \right), \nonumber
\end{eqnarray}
and
\begin{eqnarray}
 {\hat V} \equiv
V_{\alpha \beta } + \varepsilon_{\alpha}, \enskip \lambda \equiv
E. \nonumber
\end{eqnarray}
The potential matrix is the real symmetric matrix of  continuous
functions, $x \in [0, a]$. Next, we add to the equation
(\ref{schrsymb}) the following boundary conditions
\begin{eqnarray} \left\{ \begin{array}{lcl}
y'(0)-{\hat h} y(0) = 0 & & y'(a) + {\hat H} y(a) = 0 \\
y_{i}'(0)-{\hat h}_{i} y_{i}(0) = 0 & & y_{i}'(a) + {\hat H}
y_{i}(a) = 0, \enskip i=1,..., M
\end{array}
\label{bcnds} \right.
\end{eqnarray}
where we take   ${\hat h}$, ${\hat h}_{i}$ and ${\hat H}$ to all
be the real symmetric matrices.
 We denote the spectra of the M+1 problems (\ref{schrsymb}) and
(\ref{bcnds}) by $\{\lambda_{n}\}_{n=1}^{\infty}$ and
$\{\lambda_{n}^i \}_{n=1}^{\infty}$, respectively. There is no
theorem of interlacing of the spectra in M-channel case, $M > 1$.
So, we additionally require that no spectrum degeneracy should
occur.

Let us denote by ${\hat \phi}(x, \lambda)$ and ${\hat \chi}_{i}(x,
\lambda)$ the matrix solutions of the equation (\ref{schrsymb})
satisfying the initial conditions
\begin{equation}
{\hat \phi}(0, \lambda) = {\hat 1}, \enskip {\hat
\phi}'(x,\lambda)|_{x=0}={\hat h}, \enskip {\hat \chi}_{i}(0,
\lambda) = {\hat 1}, \enskip {\hat \chi}'_{i}(x, \lambda)|_{x=0} =
{\hat h}_{i}, \label{fipsi}
\end{equation}
where the prime stands for the derivative with respect to $x$. In
what follows we shall use the prime to denote this derivative
apart from the special cases the reader will be let know of.
Besides, the hat will always stand for the matrix. Do not confuse
the following: a matrix solution of (\ref{schrsymb}) means that
each column of the matrix is a vector-solution, only satisfying a
specific initial (boundary) condition.   Eigenvalues of the
boundary value problems (\ref{schrsymb}) and (\ref{bcnds})
coincide with zeros of determinants of the matrices
\begin{eqnarray} \left\{ \begin{array}{l}
{\hat \Phi}(\lambda ) = {\bar {\hat \phi}}\,'(x,\lambda)|_{x=a} +
{\bar {\hat \phi}}(a,\lambda) {\hat H} \\
{\hat \Phi }_{i}(\lambda) = {\bar {\hat
\chi}}_{i}'(x,\lambda)|_{x=a} +  {\bar {\hat \chi}}_{i}(a,\lambda)
{\hat H},
\end{array}
\label{fi} \right.
\end{eqnarray}
where the bar sign  denotes transpose.

Now we introduce the norming vectors associated with  the spectrum
$\{\lambda_{n}\}_{n=1}^{\infty}$
\begin{eqnarray}
\gamma_{\lambda _{n}} \equiv \left( \begin{array}{c} \gamma_{1}(\lambda _{n})\\
. \\
.\\
 .\\
\gamma_{M}(\lambda _{n})\end{array} \right), \nonumber
\end{eqnarray}
 so that
\begin{eqnarray}
{\hat \phi}(x, \lambda_{n}) \gamma_{\lambda _{n}} = y(x,
\lambda_{n}) \label{gamma}
\end{eqnarray}
with  $y'(x,\lambda_{n})|_{x=a} + {\hat H} y(a,\lambda_{n}) = 0$
and $\int_{0}^{a}\sum_{\alpha =1}^{M}[\Psi_{\alpha
}(x,\lambda_{n})]^2dx=1$. Likewise, for the spectra
$\{\lambda_{n}^i\}_{n=1}^{\infty}$
\begin{eqnarray}
\gamma_{\lambda _{n}^i} \equiv \left( \begin{array}{c} \gamma_{1}(\lambda _{n}^i)\\
. \\
.\\
 .\\
\gamma_{M}(\lambda _{n}^i)\end{array} \right), \nonumber
\end{eqnarray}
so that
\begin{eqnarray}
{\hat \chi}_{i}(x, \lambda_{n}^i) \gamma_{\lambda _{n}^i} =
y_{i}(x, \lambda_{n}^i) \label{gammai}
\end{eqnarray}
 with $ y_{i}'(x,\lambda_{n}^i)|_{x=a} +
{\hat H} y_{i}(a,\lambda_{n}^i) = 0$ and $\int_{0}^{a}\sum_{\alpha
=1}^{M}[\Psi_{\alpha }(x,\lambda_{n}^i)]^2dx=1$. Let us also
introduce the function $\gamma_{\lambda}$  (vs. $\lambda$) so that
$\gamma_{\lambda}=\gamma_{\lambda _{n}}$ when $\lambda = \lambda
_{n}$ and $\gamma_{\lambda}=\gamma_{\lambda _{n}^i}$ when $\lambda
= \lambda _{n}^i$. That function makes the sense at the points
$\lambda _{n}$ and $\lambda _{n}^i$ only. In between, we have the
freedom to specify it arbitrarily. We can only require this
function to be continuous differentiable and have no
singularities.

We take
\begin{eqnarray}
f_{i}(x, \lambda) \equiv  {\bar \gamma}_{\lambda} {\bar {\hat
\chi}}_{i}(x,\lambda) + m_{i}(\lambda )  {\bar \gamma}_{\lambda}
{\bar {\hat \phi}}(x,\lambda), \label{f}
\end{eqnarray}
where  $m_{i}(\lambda )$ is scalar and  we require that
\begin{eqnarray}
f_{i}'(x,\lambda)|_{x=a} + f_{i}(a,\lambda) {\hat H} = 0 \enskip
\Longrightarrow \label{fprime}
\end{eqnarray}
\begin{eqnarray}
m_{i}(\lambda )[{\bar \gamma}_{\lambda} {\bar {\hat
\phi}}\,'(x,\lambda)|_{x=a} + {\bar \gamma}_{\lambda} {\bar {\hat
\phi}}(a,\lambda) {\hat H}] = -[{\bar \gamma}_{\lambda} {\bar
{\hat \chi}}_{i}'(x,\lambda)|_{x=a} + {\bar \gamma}_{\lambda}
{\bar {\hat \chi}}_{i}(a,\lambda) {\hat H}]. \label{meq}
\end{eqnarray}
Comparing with (\ref{fi}) we have
\begin{eqnarray}
m_{i}(\lambda )=-\frac{{\bar \Phi }_{i}(\lambda) \Phi
(\lambda)}{{\bar \Phi }(\lambda) \Phi (\lambda)},
 \label{m}
\end{eqnarray}
where we denote $\Phi (\lambda ) \equiv {\bar {\hat
\Phi}}(\lambda) \gamma _{\lambda}$ and $\Phi _{i}(\lambda ) \equiv
{\bar {\hat \Phi}}_{i}(\lambda) \gamma _{\lambda}$.

Next, employing the well known Green formula we have
\begin{eqnarray}
(\lambda - \lambda _{n}) \int_{0}^{a} f_{i}(x, \lambda) {\hat
\phi}(x, \lambda_{n}) \gamma_{\lambda _{n}}dx=(\lambda - \lambda
_{n}) \int_{0}^{a} {\bar \gamma}_{\lambda} {\bar {\hat
\chi}}_{i}(x,\lambda) {\hat \phi}(x, \lambda_{n}) \gamma_{\lambda
_{n}} dx \nonumber \\
- (\lambda - \lambda _{n}) \frac{{\bar \Phi }_{i}(\lambda) \Phi
(\lambda)}{{\bar \Phi }(\lambda) \Phi (\lambda)} {\bar
\gamma}_{\lambda} {\bar {\hat \phi}}(x,\lambda) {\hat \phi}(x,
\lambda_{n}) \gamma_{\lambda _{n}}dx = f_{i}'(x,\lambda)|_{x=0}
{\hat \phi}(0, \lambda_{n}) \gamma_{\lambda _{n}} \nonumber \\  -
f_{i}(0,\lambda) {\hat \phi}(x, \lambda_{n})|_{x=0}
\gamma_{\lambda _{n}} = {\bar \gamma}_{\lambda} ({\hat h}_{i} -
{\hat h}) \gamma_{\lambda _{n}},
 \label{green}
\end{eqnarray}
where we use the definitions (\ref{f}), (\ref{fipsi}) and
(\ref{m}). The last equality follows from the fact that the
matrices in (\ref{bcnds}) are symmetric.

Let us pass to the limit $\lambda \to \lambda _{n}$. Then the
equation (\ref{green}) goes over into
\begin{eqnarray}
-\frac{\frac{d}{d \lambda} [{\bar \Phi}(\lambda ) \Phi
(\lambda)]|_{\lambda = \lambda _{n}}}{{\bar \Phi}_{i}(\lambda
_{n}) \Phi (\lambda _{n})} {\bar \gamma}_{\lambda _{n}} ({\hat
h}_{i} - {\hat h}) \gamma_{\lambda _{n}}=1,
 \label{star}
\end{eqnarray}
where we used the L'Hospital rule.

We shall prove that this formula can be represented as
\begin{eqnarray}
(\lambda_{n}^{i} - \lambda _{n})^{-1} \prod_{\mu = 1}^{\infty}\!'
\frac{\lambda_{\mu }-\lambda _{n}}{\lambda_{\mu }^{i}-\lambda
_{n}} {\bar \gamma}_{\lambda _{n}} ({\hat h}_{i} - {\hat h})
\gamma_{\lambda _{n}}=1,
 \label{ficp3}
\end{eqnarray}
where the prime denotes that we omitted, in the product, the term
with the number $n$.

Since $\Phi (\lambda )$ and $\Phi _{i}(\lambda )$ are the entire
holomorphic functions they are determined (to within constant
multipliers) by their zeros and, hence, can be represented as
follows
\begin{eqnarray}
\Phi(\lambda) = C \prod_{\mu = 1}^{\infty}(1-
\frac{\lambda}{\lambda _{\mu }}); \enskip \Phi _{i}(\lambda)=
C_{i} \prod_{\nu = 1}^{\infty}(1- \frac{\lambda}{\lambda _{\nu
}^i}).
 \label{ficc}
\end{eqnarray}
Substituting (\ref{ficc}) into (\ref{star}) we have
\begin{eqnarray}
\frac{\frac{1}{\lambda _{n}} \prod_{\mu = 1}^{\infty}\!'(1-
\frac{\lambda _{n}}{\lambda _{\mu }}) {\bar C} C} {\prod_{\nu =
1}^{\infty}(1- \frac{\lambda _{n}}{\lambda _{\nu }^i}) {\bar
C}_{i} C} {\bar \gamma}_{\lambda _{n}} ({\hat h}_{i} - {\hat h})
\gamma_{\lambda _{n}}=1.
 \label{ficp}
\end{eqnarray}
Now we have to ascertain the expression for the ${\bar C} C /{\bar
C}_{i} C$. We shall need some knowledge about an asymptotic
behaviour of the solutions of (\ref{schrsymb}). First of all,
these equations become uncoupled in the limit $\lambda \to
\infty$. So, as in one-channel case, we have $\lim_{\lambda \to
\infty}{\hat \Phi}(\lambda )\{{\hat \Phi}_{i}(\lambda )\}^{-1}=1$,
and the same for the transpose of these matrices. Taking this into
account  we obtain

\begin{eqnarray}
\frac{{\bar C} C}{{\bar C}_{i} C} \prod_{\mu = 1}^{\infty}
\frac{\lambda _{\mu }^{i}}{\lambda _{\mu }}\lim_{\lambda \to
\infty}\prod_{\mu = 1}^{\infty} \frac{\lambda_{\mu
}-\lambda}{\lambda_{\mu}^{i}-\lambda}=1.
 \label{ficp1}
\end{eqnarray}

We have the following  asymptotic formulas for $\lambda $ and
$\lambda^{i}$: $\lambda_{\mu} = \mu ^2 + O(1)$ and the same for
$\lambda ^{i}$. Then $\lambda_{\mu}^{i}-\lambda_{\mu}=O(1)$ and
the series $\sum_{\mu}^{\infty}|(\lambda_{\mu}-
\lambda_{\mu}^{i})/(\lambda_{\mu}^{i} - \lambda)|$
 converges
uniformly as $\lambda \to \infty$. Hence, we can pass to the limit
in each term of the infinite product
\begin{eqnarray} \lim_{\lambda \to \infty}\prod_{\mu =
1}^{\infty} \frac{\lambda_{\mu
}-\lambda}{\lambda_{\mu}^{i}-\lambda}= \lim_{\lambda \to \infty}
\prod_{\mu = 1}^{\infty}(1+\frac{\lambda_{\mu}-
\lambda_{\mu}^{i}}{\lambda_{\mu}^{i} - \lambda})=1.
 \label{infprod}
\end{eqnarray}
We see from (\ref{infprod}) and (\ref{ficp1}) that

\begin{eqnarray}
\frac{{\bar C} C}{{\bar C}_{i} C} \prod_{\mu = 1}^{\infty}
\frac{\lambda _{\mu }^{i}}{\lambda _{\mu }}=1.
 \label{ficp2}
\end{eqnarray}

At last, we can obtain the final expression for $\gamma
_{\lambda_{n}}$. Substituting (\ref{ficp2}) into (\ref{ficp}) we
have the formula (\ref{ficp3})-- the system of $M$ equations
($i=1,...,M$) for determining $M$ components of $\gamma_{\lambda
_{n}}$.

In the one-channel case the formula (\ref{ficp3}) goes over into
the  known expression for two spectra:
\begin{eqnarray}
(\lambda_{n}^{2} - \lambda _{n}^{1})^{-1} \prod_{\mu =
1}^{\infty}\!' \frac{\lambda_{\mu }^{1}-\lambda
_{n}^{1}}{\lambda_{\mu }^{2}-\lambda _{n}^{1}} (h_{2} - h_{1})
\gamma_{\lambda _{n}}^2=1,
 \label{ficp2sp}
\end{eqnarray}
where the matrix values become scalars, and we denote, by indeces
1 and 2, two spectra determining scalar norming factor
$\gamma_{\lambda _{n}}$.

The system (\ref{ficp3}) is not linear one: each row in it
contains the quadratic form ${\bar \gamma}_{\lambda _{n}} ({\hat
h}_{i} - {\hat h}) \gamma_{\lambda _{n}}$. Hence, these equations
cannot be solved uniquely in general (including solvability
itself). In other words, we have to impose some constraint on
choosing the matrices ${\hat h}_{i}$, i.e. the difference ${\hat
h}_{i} - {\hat h}$. Among other possibilities, we give several
realizations which will allow a unique solvability of the system
(\ref{ficp3}).

\bigskip

i) The symmetric matrix ${\hat h}_{i} - {\hat h} \equiv {\hat
\xi}^{(i)}$ has the form of a Jacobi matrix:
\begin{equation}
{\hat \xi}^{(i)}= \left( \begin{array}{ccccccc}  \xi_{11}^{(i)} &
\xi_{12}^{(i)} & 0 & 0 & . & .  & 0\\ \xi_{12}^{(i)} & 0 &
\xi_{23}^{(i)} & 0 & .  & . & .
\\ 0 & \xi_{23}^{(i)} & 0 & \xi_{34}^{(i)} & . & . & .\\ .  & .  & . &
. & . & . & .\\ . & . & . & .  & . & . & \xi_{M-1 M}^{(i)} \\ 0 &
. & . & . & 0 & \xi_{M-1 M}^{(i)} & 0 \end{array} \right),
\label{xi1}
\end{equation}
where the main diagonal contains only one non-zero element,
$\xi_{11}^{(i)}$. Then
\begin{equation}
{\bar \gamma}_{\lambda _{n}} ({\hat h}_{i} - {\hat h})
\gamma_{\lambda _{n}}=\xi_{11}^{(i)} \gamma_{1}(\lambda _{n})^2 +
2 \sum_{k \ne 1 }^{M} \xi_{k-1 k}^{(i)} \gamma_{k-1}(\lambda _{n})
\gamma_{k}(\lambda _{n}). \label{xi2}
\end{equation}
Introducing the variables $\omega_{1} \equiv \gamma_{1}(\lambda
_{n})^2$ and $\omega_{k} \equiv \gamma_{k-1}(\lambda _{n})
\gamma_{k}(\lambda _{n}), \enskip k=2,...M$ we can rewrite the
last expression as follows

\begin{equation}
{\bar \gamma}_{\lambda _{n}} ({\hat h}_{i} - {\hat h})
\gamma_{\lambda _{n}}=\xi_{11}^{(i)} \omega_{1} + 2 \sum_{k \ne 1
}^{M} \xi_{k-1 k}^{(i)} \omega_{k}. \label{xi3}
\end{equation}

Then (\ref{ficp3}) becomes the system of linear algebraic
equations for the variables $\omega$. If $\omega_{1} =
\gamma_{1}(\lambda _{n})^2 >0$, then $\gamma_{1}(\lambda _{n})=\pm
\omega_{1}^{1/2}$, $\gamma_{2}(\lambda _{n})=\mp
\omega_{2}/\omega_{1}^{1/2}$ and so forth. The sign in front of
$\omega_{1}^{1/2}$ in the expression for $\gamma_{1}(\lambda
_{n})$ determines the common sign for $\gamma_{\lambda _{n}}$ and,
hence, is inessential: The whole vector-valued wave function is
determined to within  sign ($\pm$). With the non-zero element
$\xi_{ll}^{(i)} \ne 0, \enskip l \ne 1$ positioned in arbitrary
place of the main diagonal, the scheme is analogous.

\bigskip

ii) The matrix ${\hat h}_{i} - {\hat h} \equiv {\hat \zeta}^{(i)}$
is represented as follows:
\begin{equation}
{\hat \zeta}^{(i)}= \left( \begin{array}{ccccccc}
0 & . & 0 & \zeta_{1l}^{(i)} & 0 & .  & 0\\
. & . & . & . & .  & . & .
\\ . & . & 0 & \zeta_{l-1l}^{(i)} & 0 & . & .\\ \zeta_{l1}^{(i)}  & .  & \zeta_{ll-1}^{(i)} &
\zeta_{ll}^{(i)} &  \zeta_{ll+1}^{(i)}& . & \zeta_{lM}^{(i)}\\ . &
. & 0 & \zeta_{l+1l}^{(i)}  & 0 & . & 0 \\
. & . & . & . & . & . & . \\ 0 & . & 0 & \zeta_{Ml}^{(i)} & 0 & .
& 0
\end{array} \right), \label{zeta1}
\end{equation}
i.e., the matrix contains one non-zero row and one non-zero column
which cross each other in a place of the entry $\zeta_{ll}^{(i)}$.
For the quadratic form we have (using the symmetry of ${\hat
h}_{i} - {\hat h}$)
\begin{equation}
{\bar \gamma}_{\lambda _{n}} ({\hat h}_{i} - {\hat h})
\gamma_{\lambda _{n}}=\zeta_{ll}^{(i)} \gamma_{l}(\lambda _{n})^2
+ 2 \sum_{k \ne l }^{M} \zeta_{l k}^{(i)} \gamma_{l}(\lambda _{n})
\gamma_{k}(\lambda _{n}). \label{zeta2}
\end{equation}
Introducing new variables $\theta _{k}\equiv \gamma_{l}(\lambda
_{n}) \gamma_{k}(\lambda _{n}), \enskip k \ne l$ and $\theta _{l}
\equiv  \gamma_{l}(\lambda _{n})^2$ we can now look upon
(\ref{ficp3}) as a linearized system again:
\begin{eqnarray}
(\lambda_{n}^{i} - \lambda _{n})^{-1} \prod_{\mu = 1}^{\infty}\!'
\frac{\lambda_{\mu }-\lambda _{n}}{\lambda_{\mu }^{i}-\lambda
_{n}} \{\zeta_{ll}^{(i)} \theta_{l} + 2 \sum_{k \ne l }^{M}
\zeta_{l k}^{(i)} \theta _{k}\}=1. \label{zeta3}
\end{eqnarray}
After deriving $\theta _{i}$, one can obtain $\gamma_{i}(\lambda
_{n})$ trivially. Of course, the solvability in this case depends
on whether the corresponding determinant for the system
(\ref{zeta3}) is non-zero and $\theta _{l} >0$.

In all the cases, the knowledge of the complete set
$\{\lambda_{n},\gamma_{\lambda _{n}} \}_{n=1}^{\infty}$ allows a
unique restoration of the potential matrix by the standard
Gel'fand-Levitan theory (its multichannel generalization).

\section{Conclusions}

In this paper, the relationship is established  between components
of the norming vector $ \gamma_{\lambda _{n}}$ associated with a
certain boundary value problem (with the spectrum $\{\lambda _{n}
\}_{n=1}^{\infty}$) and the spectra (including $\{\lambda _{n}
\}_{n=1}^{\infty}$) of $M+1$  multichannel Sturm-Liouville
operators with the same potential matrix $V_{\alpha \beta}(x)$ but
different boundary conditions. As a matter of fact, the central
result is the formula (\ref{ficp3}). Though giving no unique
solutions in general, it can get linear if we require the matrices
${\hat h}_{i}$ to be of special type. Hence, the uniqueness of the
multichannel inverse eigenvalue problem from $M+1$ spectra  is
however possible for a particular class of boundary conditions.
The problem of specifying the necessary and sufficient conditions
needs a special examination. It is clear that scrutinizing the
asymptotic behaviour of the spectra with different boundary
conditions will be required. It is closely associated with
specifying the class of differentiable functions the $V_{\alpha
\beta}(x)$ pertain to. So, the results given present only an
intermediate stage in investigations on the subject.


\begin{thebibliography}{99
}
\bibitem{F} Feshbach H 1962 {\it Ann.Phys., NY} {\bf 19} 287-313.
\bibitem{ZS} Zakhariev B N and Suzko A A 1990 {\it Direct and Inverse
 Problems} (Heidelberg: Springer)
\bibitem{Cox} Cox J R 1966 {\it Ann.Phys., NY} {\bf 39} 216-236; 1967 {\it
J. Math. Phys.} {\bf 8} 2327-31.
\bibitem{CoxGar} Cox J R and Garcia H R 1975 {\it J. Math. Phys.}
{\bf 16} 1402-09.
\bibitem{ChSab} Chadan K and Sabatier P 1989 {\it Inverse Problems in
Quantum Scattering Theory} 2nd edn (Heidelberg: Springer)
\bibitem{Borg} Borg G 1946 {\it Acta Math.} {\bf 78} 1-96.
\bibitem{Lev} Levitan B M 1987 {\it Inverse Sturm-Liouville Problems}
(Zeist, The Netherlands: VSP).
\bibitem{Zak} Zakhariev B N 1990  {\it JINR Rapid
Communications} N6[45]-90 41-48.

  \end{thebibliography}
\end{document}